\documentclass[aps,prb,showpacs,twocolumn,amsmath,amssymb,superscriptaddress]{revtex4-1}

\usepackage{graphicx}
\usepackage{subfigure}
\usepackage{grffile}
\usepackage{bm}
\usepackage{mhchem}
\usepackage{color}
\usepackage{hyperref}

\newcommand{\ket}[1]{{\vert #1\rangle}}

\newcommand{\eval}[3]{\langle#1\vert#2\vert#3\rangle}
\newcommand{\vev}[1]{\langle #1\rangle}
\newcommand{\1}{\mbox{\bf 1}}
\newcommand{\ud}{\mathrm{d}}
\newcommand{\RE}{\text{Re}}

\begin{document}

\title{Numerical method to compute optical conductivity based on pump-probe simulations}

\author{Can Shao}
\affiliation{Center for Interdisciplinary Studies $\&$ Key Laboratory for Magnetism and Magnetic Materials of the MoE, Lanzhou University, Lanzhou 730000, China}

\author{Takami Tohyama}
\affiliation{Department of Applied Physics, Tokyo University of Science, Tokyo 125-8585, Japan}

\author{Hong-Gang Luo}
\affiliation{Center for Interdisciplinary Studies $\&$ Key Laboratory for Magnetism and Magnetic Materials of the MoE, Lanzhou University, Lanzhou 730000, China}
\affiliation{Beijing Computational Science Research Center, Beijing 100084, China}

\author{Hantao Lu}
\email{luht@lzu.edu.cn}
\affiliation{Center for Interdisciplinary Studies $\&$ Key Laboratory for Magnetism and Magnetic Materials of the MoE, Lanzhou University, Lanzhou 730000, China}

\date{\today}

\begin{abstract}
A numerical method to calculate optical conductivity based on a pump-probe setup is presented. Its validity and limits are tested and demonstrated via concrete numerical simulations on the half-filled one-dimensional extended Hubbard model both in and out of equilibrium. By employing either a steplike or a Gaussian-like probing vector potential, it is found that in nonequilibrium, the method in the narrow-probe-pulse limit can be identified with variant types of linear response theory, which, in equilibrium, produce identical results. The observation reveals the underlying probe-pulse dependence of the optical conductivity calculations in nonequilibrium, which may have applications in the theoretical analysis of ultrafast spectroscopy measurements.
\end{abstract}


\maketitle

\section{Introduction}
\label{sec1}

Time-resolved spectroscopy has been employed extensively to investigate the dynamic properties of materials with strong correlations, e.g., (quasi-)one-dimensional (1D) Mott insulators such as bis(ethylendithyo)-tetrathiafulvalene-difluorotetracyanoquinodimethane ET-F$_2$TCNQ~\cite{Okamoto07,Wall11,Mitrano14} and halogen-bridged transition-metal compounds ~\cite{Iwai03,Matsuzaki14}, as well as correlated systems with higher space dimensions, such as cuprates~\cite{Fausti11,Nicoletti14,Hu14,Contei15} and manganites~\cite{Rini07,Li13}. As a promising time-resolved spectroscopy technique, ultrafast pump-probe optical measurement is able to unravel the complicated entanglement of various degrees of freedom in a correlated system to some extent by resolving their contributions separately at multiple time scales~\cite{Orenstein12}. For experiments on electronic systems, time-resolved optical conductivity is an essential quantity, from which the dynamic properties of charges and other related degrees of freedom can be analyzed; see the discussions in Refs.~\onlinecite{Orenstein15,Nicoletti15}.

Generally, in order to calculate a dynamic response function, the corresponding correlation functions at different times should be obtained. In nonequilibrium, the task becomes more demanding since the system is evolving and the time-translation invariance is absent. Consequently, the correlation functions with two time variables cannot be reduced to functions with only one time variable, i.e., the temporal distance in the case of equilibrium. Theoretically, there are various ways in microscopic model calculations to describe the dynamic response of correlated systems in nonequilibrium. One possible way is based on the nonequilibrium Green's functions: by using the Kadanoff-Baym or Keldysh formalism, methods that have been applied to correlated systems in equilibrium, such as dynamic mean-field theory, can be adapted to nonequilibrium on a formal level (see, e.g., Ref.~\onlinecite{Aoki14} and references therein). Another method begins with wave functions; this method is used primarily within closed systems. The time-dependent wave functions can be secured using various numerical methods, such as exact diagonalization and the density-matrix renormalization-group method. Thereafter, the time-dependent expectation values of observables and various temporal correlations can be readily obtained, which enables further calculations on the out-of-equilibrium dynamic response. Specifically, with regard to time-dependent optical conductivity, several related but different schemes have been employed in similar models, including the nonequilibrium generalizations of the Kubo formula~\cite{Filippis12}, linear-response theory~\cite{Rossini14,Zala14}, or a direct calculation on dynamic current-current correlations for transient states~\cite{Fukaya15}. However, the underlying characters and validity of these approaches have not yet been fully addressed as far as we know. This is also a practical problem for the purpose of analyzing data from (sub-)THz spectroscopy measurements, where the temporal resolution has been pushed up to the order of a few tens of femtoseconds~\cite{Contei15}.

In this paper, we suggest that the way in which temporal correlations are incorporated can be crucial in the study of nonequilibrium dynamics. Inspired by the pump-probe setup in the experiments, we propose a method to calculate the optical conductivity. Its validity and limits are demonstrated via detailed numerical simulations on the 1D half-filled extended Hubbard model, a prototype of strongly correlated electronic systems. A comparison between our method and the existing ones shows that by adopting two different limits on the probe pulse, our method can reproduce the results of two types of linear-response theory. We note that in some nonequilibrium situations, the two sets of results on the time-dependent optical conductivity can be quite different. Theoretical analysis combined with numerical simulations elucidates the difference and connections between these various methods to a satisfactory level. They also raise the issue of the probe-pulse dependence in nonequilibrium analysis, which has been ignored in previous studies.

The rest of the paper is organized as follows. In Sec.~\ref{sec2}, after the model description and a critical review of the relevant existing methods, our approach is presented. The following section provides a theoretical analysis on the validity of the approach, with numerical demonstrations concerning the options for probe-pulse parameters. Section~\ref{sec4} contains comprehensive comparisons and analysis. The conclusion is drawn in Sec.~\ref{sec5}.

\section{Model and methods}\label{sec2}

The Hamiltonian we use in the discussion is the well-known 1D extended Hubbard model at half-filling:
\begin{eqnarray}
H&=&-t_h\sum_{i,\sigma}\left(c_{i,\sigma}^{\dagger}
c_{i+1,\sigma}+\text{H.c.}\right) \nonumber \\
&+&U\sum_{i}\left(n_{i,\uparrow}-\frac{1}{2}\right)
\left(n_{i,\downarrow}-\frac{1}{2}\right) \nonumber \\
&+&V\sum_i\left(n_i-1\right)\left(n_{i+1}-1\right),
\label{eq:1}
\end{eqnarray}
where $c_{i,\sigma}^\dagger$ ($c_{i,\sigma}$) is the creation (annihilation) operator of an electron with spin $\sigma$ at site $i$, $n_{i,\sigma}=c_{i \sigma}^\dagger c_{i,\sigma}$, $n_i=n_{i,\uparrow}+n_{i,\downarrow}$, $t_h$ is the hopping constant, and $U$ and $V$ are on-site and nearest-neighbor Coulomb repulsion interactions, respectively. In the following, we set $t_h$ and $t_h^{-1}$ as energy and time units and take the Plank constant $\hbar=1$, and the elementary charge $e=1$.

We restrict our discussions to zero temperature. Before presenting our method in detail, we first present some existing methods that are relevant to our discussions.

In the linear-response theory, the induced current due to the application of an external perturbation field $E(t)$ is given by (in the 1D case)
\begin{equation}
\vev{j(t')}=\int_{-\infty}^{t'}\sigma(t',t)E(t)\,\ud t,
\label{eq:5}
\end{equation}
where the {\em response function} $\sigma(t',t)$, which is probe-field-independent, measures the current response at time $t'$ with respect to the perturbation of a {\em unit pulse} at time $t$~\cite{Kubo57}. After the response function is determined, the induced current for any given perturbation field $E(t)$ in the linear-response regime can be calculated accordingly by Eq.~(\ref{eq:5}).

The situation can be simplified in equilibrium, where due to the presence of time-translation invariance (without consideration of the probe field), the two-time response function $\sigma(t',t)$ can be reduced to a single-variable function as $\sigma(t'-t)$. Then the Fourier transformation on Eq.~(\ref{eq:5}) simply produces
\begin{equation}
	j(\omega)=\sigma(\omega)E(\omega).
	\label{eq:2.5}
\end{equation}
$\sigma(\omega)$ represents the optical conductivity, which measures the current response of the system in {\em frequency space}. It is worth noting that although the conductivity $\sigma(\omega)$ can be evaluated or measured by $j(\omega)/E(\omega)$ as indicated in Eq.~(\ref{eq:2.5}) (see additional discussions in Sec.~(\ref{sec3})), it reflects the property of the system in equilibrium and does not depend on the details of the probe field. It is well known that $\sigma(\omega)$ can be calculated microscopically by the Kubo formula (KF)~\cite{Kubo57} in terms of the current-current correlations. The regular part (excluding the possible singularity at $\omega=0$) is given by (e.g., in Ref.~\onlinecite{Mahan13}):
\begin{equation}
	\sigma_{\text{reg}}(\omega)=\frac{1}{\omega L}\int_0^{+\infty}e^{i\omega s}\eval{0}{[j(s),j(0)]}{0}\,\ud s=\frac{i}{\omega L} \chi(\omega),
	\label{eq:2}
\end{equation}
where $\ket{0}$ is the ground state of the system, $L$ is the lattice size, and the electric susceptibility $\chi(\omega)$ is defined as
\begin{equation}
	\chi(\omega)=\int_{-\infty}^{+\infty}e^{i\omega{s}}(-i)\theta(s)\eval{0}{[j(s),j(0)]}{0}\,\ud s.
\label{eq:3}
\end{equation}
The current $j(s)$ in Eqs.~(\ref{eq:2}) and (\ref{eq:3}) is the Heisenberg representation of the current operator $j$, which reads
\begin{equation}
	j=-it_h\sum_{i,\sigma}[c_{i,\sigma}^{\dagger}c_{i+1,\sigma}-\text{H.c.}].
\label{eq:4}
\end{equation}

In nonequilibrium, on the other hand, since time-translation invariance does not hold in general, a full two-time response function is required in order to describe the linear response of the electric current with respect to a time-dependent probing field $E(t)$. However, for another related and important quantity, i.e., the time-dependent optical conductivity that we will mainly focus on, there is no unique definition if the time-translation invariance does not hold in the response function~\cite{Eckstein10,Eckstein13,Zala14}. Following Ref.~\onlinecite{Zala14}, throughout the article we define the conductivity as:
\begin{equation}
\sigma(\omega,t)=\int_0^{t_m}\sigma(t+s,t)e^{i(\omega+i\eta)s}\,\ud s,
\label{eq:6}
\end{equation}
where we have explicitly introduced a spectral broadening factor $\eta$ and the time width $t_m$ for the Fourier transformations that will be used in the numerical simulations. In practice, $t_m$ is the maximum of the probe duration time, which in THz spectroscopy is usually several hundred times larger than the microscopic time unit (e.g., $t_h^{-1}$ here).

From Eq.~(\ref{eq:6}), it is clear that if we intend to calculate $\sigma(\omega,t)$ by investigating the current response to a weak probe field analogous to Eq.~(\ref{eq:2.5}), which will be developed as the pump-probe method later in the article, a highly narrow probe pulse is preferred. However, different from the equilibrium case, we find that the response character of the system in nonequilibrium depends on the form of the probe pulse that we choose, which will be detailed in Sec.~\ref{sec4}. Here, we first review the relevant theoretical formulas related to $\sigma(\omega,t)$.

In the linear-response regime, rigorous analysis produces the response function $\sigma(t',t)$ as (with restriction $t'\ge t$)~\cite{Rossini14,Zala14}
\begin{equation}
\sigma(t',t)=\frac{1}{L}\left[\eval{\psi(t')}{\tau}{\psi(t')}+\int_t^{t'}\chi(t',t'')\,\ud t''\right],
\label{eq:8}
\end{equation}
where the two-time susceptibility is
\begin{equation}
	\chi(t',t'')=-i\theta(t'-t'')\eval{\psi(t)}{[j^{I}(t'),j^{I}(t'')]}{\psi(t)},
\label{eq:9}
\end{equation}
and in the diamagnetic term, $\tau=t_h\sum\limits_{i,\sigma}(c_{i+1,\sigma}^{\dagger}c_{i,\sigma}+\text{H.c.})$, which is the stress tensor operator. The interaction representation of an operator $B^{I}(t')$ is defined as $U^\dagger(t',t)\ B\ U(t',t)$, where $U(t',t)$ is the time-evolution operator {\em in the absence of probing perturbations}~\cite{Zala14}. Consistent with this, the time-dependent wave function $\ket{\psi(t)}$ follows the nonequilibrium time evolution of the (closed) system without the probe field. It is clear from the expression that in order to obtain the value of $\sigma(t',t)$, the current-current correlations (commutators) between the intermediate time and the ending time $t'$ have to be taken into account. We refer to this as nonequilibrium linear response (NLR). In the next section, we will show that the Fourier transformation of $\sigma(t',t)$ according to the definition~(\ref{eq:6}) reflects the current response with respect to the unit electric pulse at time $t$.

\begin{figure}
\centering
\includegraphics[width=0.45\textwidth]{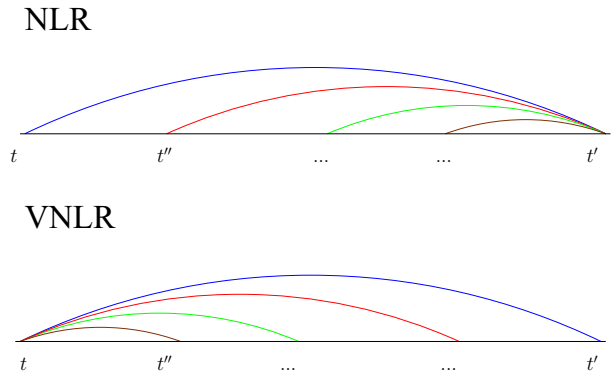}
\caption{(Color online) A schematic diagram of the current-current correlations considered in NLR and VNLR, respectively.}
\label{fig1}
\end{figure}

In the literature, we notice that another kind of generalized Kubo formula in nonequilibrium (NKF) has been used~\cite{Filippis12, Wall11}:
\begin{align}
	\text{Re}\,\sigma_{\text{reg}}(\omega,t)=&\frac{1}{\omega{L}}\text{Im}\,\int_0^{t_m}\,\ud s\ \left[ie^{i(\omega+i\eta)s}\right.\nonumber\\
													 \times&\left.\eval{\psi(t)}{[j^{I}(t+s),j^{I}(t)]}{\psi(t)}\right].
\label{eq:7}
\end{align}
A careful analysis shows that the NKF cannot be obtained from the response function in Eq.~(\ref{eq:8}) (with the diamagnetic term dropped) by using the definition of Eq.~(\ref{eq:6}). It seems that another "response function", denoted as $\tilde{\sigma}(t',t)$ $(t'\ge t)$ here, is used:
\begin{equation}
	\tilde{\sigma}(t',t)=\frac{1}{L}\left[\eval{\psi(t')}{\tau}{\psi(t')}+\int_t^{t'} \chi(t'',t)\,\ud t''\right],
\label{eq:10}
\end{equation}
where similarly
\begin{equation}
	\chi(t'',t)=-i\theta(t''-t)\eval{\psi(t)}{[j^{I}(t''),j^{I}(t)]}{\psi(t)}.
\label{eq:11}
\end{equation}
We refer to this as a variant of NLR (VNLR). First, it is easy to recognize that the NKF in Eq.~(\ref{eq:7}) is just the real regular part of the Fourier transformation of the second term in Eq.~(\ref{eq:10}) with respect to $t'-t$. Furthermore, the difference between NLR and VNLR actually comes from the second term in the response functions (Eqs.~(\ref{eq:8}) and (\ref{eq:10})), where under the integration over the intermediate time $t''\in(t,t')$, the two-time commutators of the current operator take place at $t''$ to $t'$, and $t$ to $t''$, respectively. Note that $t$ and $t'$ are the starting time and ending time of $\sigma(t',t)$ ($t'\ge t$), respectively. That is to say, different from NLR, $\tilde{\sigma}(t',t)$ takes into account the current-current correlations (commutators) between the starting time $t$ and the intermediate time. The above description can be read from Fig.~\ref{fig1}. In the next section, we will show that the Fourier transformation of $\tilde{\sigma}(t',t)$ according to the definition~(\ref{eq:6}) reflects the current response with respect to a delta vector potential applied at time $t$.

We argue that the NLR method, which has been derived rigorously from linear-response theory~\cite{Zala14}, contains more complete information about the temporal correlations than VNLR (or equivalently NKF) {\em with respect to the definition of $\sigma(\omega,t)$ in Eq.~(\ref{eq:6})}. The observation can be easily seen in Fig.~\ref{fig1}, reminding us that $\sigma(\omega,t)$ is the Fourier transformation of $\sigma(t',t)$ with respect to $t'-t$ (with $t$ fixed). The fact suggests that VNLR can be regarded as a kind of partial summation on the temporal correlations of NLR. It should be emphasized that for the equilibrium case, the two produce identical $\sigma(\omega)$ due to the time-translation invariance. Then an issue could be raised regarding the validity and limitation of VNLR: in what situation and to what extend can VNLR remain a good approximation? Does it impose any specific restriction on the probing fields? To address this issue, we now present our method.


Inspired by the experimental pump-probe setup and the theoretical works on the coherent dynamics in the BCS model~\cite{Papenkort07,Schnyder11}, we propose a method to calculate the optical conductivity of the Hubbard model both in and out of equilibrium as follows.

First, one notes that an external spatially homogeneous electric field applied along the chain in the Hamiltonian~(\ref{eq:1}) can be incorporated via the Peierls substitution in the hopping terms as~\cite{Denis12}:
\begin{equation}
c^{\dagger}_{i,\sigma}c_{i+1,\sigma}\rightarrow
e^{iA(t)}c^{\dagger}_{i,\sigma}c_{i+1,\sigma},
\qquad
H\rightarrow H(t).
\label{eq:12}
\end{equation}
With the knowledge of the wave function $\psi(t)$ under the action of $A(t)$, the temporal evolution of the expectation value of the current operator $\vev{j(t)}$, defined as $\vev{j(t)}=\eval{\psi(t)}{j(t)}{\psi(t)}$, can be readily obtained. Note that with the existence of $A(t)$, the current operator $j$ becomes time dependent:
\begin{equation}
	j(t)=\frac{\delta H(t)}{\delta A(t)}=-it_h\sum_{i,\sigma}[e^{iA(t)}c_{i,\sigma}^{\dagger}c_{i+1,\sigma}-\text{H.c.}].
\label{eq:13}
\end{equation}

In equilibrium, we set the vector potential $A(t)$ to be a weak probe pulse denoted as $A_{\text{probe}}(t)$, and the current induced by it as $\vev{j_{\text{probe}}(t)}$. Here the time-dependent wave function $\ket{\psi(t)}$ describes the time evolution of the system under the influence of $A_{\text{probe}}(t)$ starting from the ground state. Identical to Eq.~(\ref{eq:2.5}), the optical conductivity can then be calculated by
\begin{equation}
	 \sigma(\omega)=\frac{j_{\text{probe}}(\omega)/L}{E_{\text{probe}}(\omega)}=\frac{j_{\text{probe}}(\omega)}{i{(\omega+i\eta)}LA_{\text{probe}}(\omega)},
\label{eq:14}
\end{equation}
where $j_{\text{probe}}(\omega)$ and $A_{\text{probe}}(\omega)$ are the Fourier transformations of $\vev{j_{\text{probe}}(t)}$ and $A_{\text{probe}}(t)$, respectively. Note that numerically, a damping factor $e^{-\eta t}$ is also introduced when the Fourier transformations are performed, as indicated in Eq.~(\ref{eq:6}). The same $\eta$ in the denominator of Eq.~(\ref{eq:14}) is necessary to distinguish the Drude component in the spectral weight at $\omega\to 0$, which will be discussed later in more detail in Sec.~\ref{sec4}.

It is worth to stress here that Eq.~(\ref{eq:14}) can be regarded as a practical definition of the conductivity in equilibrium: $\sigma(\omega)$ obtained in this way does not depend on the details of the probing field we choose, which has been verified in the numerical simulations. We are now in the position to generalize the algorithm to nonequilibrium, where the system can be driven either by a pump pulse or a quench. The key is that in order to identify the response of the system with respect to the later probe pulse, a subtraction is necessary, i.e., two successive steps are involved (each step follows a similar process to that described above) in order to calculate the optical conductivity in nonequilibrium~\cite{Lu15}. First, a time-evolution process that describes the nonequilibrium development of the system in the absence of a probe pulse is evaluated, and we obtain $\vev{j(t)}$. Secondly, in the presence of an additional probe pulse, we get $\vev{j_{\text{total}}(t)}$. The subtraction of $\vev{j(t)}$ from $\vev{j_{\text{total}}(t)}$ produces the required $\vev{j_{\text{probe}}(t)}$, i.e., the variation of the current expectations due to the presence of the probe pulse. Then analogous to Eq.~(\ref{eq:14}), the optical conductivity in nonequilibrium is proposed to be
\begin{equation}
\sigma(\omega,t_{\text{probe}})=\frac{{j_{\text{probe}}}(\omega,t_{\text{probe}})}{i{(\omega+i\eta)}LA_{\text{probe}}(\omega)},
\label{eq:15}
\end{equation}
where $t_{\text{probe}}$ is the probing time. This pump-probe-based method is abbreviated as PP. The main reason for considering the PP method seriously is that for us it seems to be a reasonable numerical approach to the ultrafast spectroscopy experiments.

A few final words are in order regarding the numerical method: to trace the temporal evolution of the system, we employ the time-dependent Lanczos method to evaluate $|\psi(t)\rangle$. The key formula is
\begin{equation}
|\psi(t+\delta{t})\rangle\simeq\sum_{l=1}^{M}{e^{-i\epsilon_l\delta{t}}}|\phi_l\rangle\langle\phi_l|\psi(t)\rangle,
\label{eq:16}
\end{equation}
where $\epsilon_l$ and $|\phi_l\rangle$ are eigenvalues and eigenvectors of the tridiagonal matrix generated in the Lanczos iteration, respectively, $M$ is the dimension of the Lanczos basis, and $\delta{t}$ is the minimum time step. More details of the algorithm can be found in Ref.~\onlinecite{Prelovsekbook}.

\section{The influence of the probe pulses - a theoretical argument} \label{sec3}

In the PP method, analogous to experiments, it is important that the strength of the probe pulse should be weak enough not to disturb the system's dynamics qualitatively. Meanwhile, the influence of the forms of the probe pulse is also crucial in nonequilibrium. This point will be made clear in the following arguments.

We begin with the specific form of Eq.~(\ref{eq:5}) in 1D:
\begin{equation}
	\vev{j_{\text{probe}}(t')}=\int_{t_0}^{t'}\sigma(t',t'')E_{\text{probe}}(t'')\,\ud t'',
\label{eq:17}
\end{equation}
where we have assumed that the probe electric field is turned on after $t_0$ with the restriction of $t'>t_0$. The Fourier transformation on $\vev{j_{\text{probe}}(t')}$ gives
\begin{align}
&j_{\text{probe}}(\omega)={\int_{t_0}^{t_M}} \vev{j_{\text{probe}}(t')}e^{i{\omega}t'}\,\ud t' \nonumber \\
=&{\int_{t_0}^{t_M}}\int_{t_0}^{t'}\ \sigma(t',t'')E_{\text{probe}}(t'')e^{i\omega t'}\,\ud t''\,\ud t' \nonumber \\
=&{\int_{t_0}^{t_M}}\left[\int_{t''}^{t_M}\sigma(t',t'')
e^{i\omega (t'-t'')}\,\ud t'\right] E_{\text{probe}}(t'')e^{i\omega t''}\,\ud t''\nonumber  \\
=&{\int_{t_0}^{t_M}}\sigma(\omega,t'')E_{\text{probe}}(t'')e^{i\omega t''}\,\ud t'',
\label{eq:18}
\end{align}
where $t_M$ is the maximum evolution time and $t_M\gg t_0$. We can see that in order to obtain $\sigma(\omega,t)$ [assuming $t\in(t_0,t_M)$, and in this section we use $t$ to denote the probing time], the most convenient choice of $E_{\text{probe}}$ in theory is a $\delta$ pulse with the form of $E_{\text{probe}}(t'')\sim \delta(t''-t)$, and the desired $\sigma(\omega,t)$ is simply proportional to $j_{\text{probe}}(\omega)$. The strategy is different from the usual treatment for equilibrium in the textbooks, e.g., in Ref.~\onlinecite{Mahan13}, where a monochromatic perturbation field $E(t)\sim e^{-i\omega t}$ is applied from $t\to-\infty$. In the temporal gauge, the favorable $\delta$ electric field corresponds to a step-like vector potential, i.e., $A_{\text{probe}}(t'')=A_{0,\text{step}}\,\theta(t''-t)$, where $\theta(t'')$ is the Heaviside step function (the effect of ramping, i.e., the gradual increase of $A_{\text{probe}}$ from zero to a finite value, instead of a jump, will be discussed later in this section). We call this kind of PP way with a step-like vector potential a PP-step. From Eq.~(\ref{eq:18}) it is easy to convince oneself that under the step potential, Eq.~(\ref{eq:15}) exactly produces $\sigma(\omega,t)$ in NLR. This means that by applying step-like vector potentials at time $t$, we can extract the exact value of $\sigma(\omega,t)$ by simply looking at the time evolution of the induced current $\vev{j_{\text{probe}}(t)}$. Thus the connection between PP-step and NLR can be established. This proposition will be verified by later numerical simulations.

\begin{figure}
\centering
\includegraphics[width=0.47\textwidth]{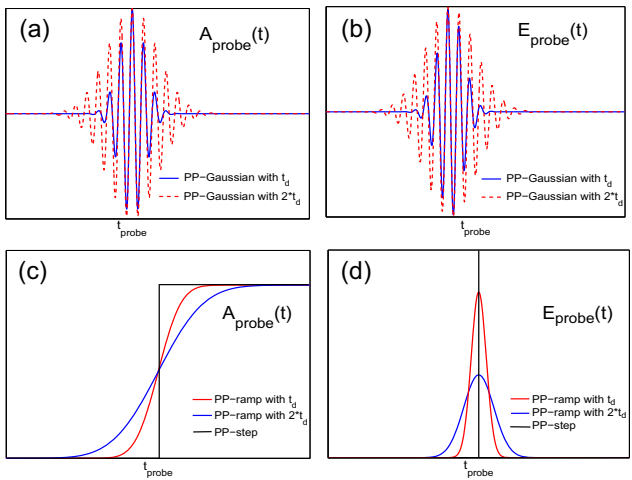}
\caption{(Color online) Schematic illustrations of the vector potentials and the corresponding electric fields for the Gaussian pulse [(a) and (b)] and the steplike pulse [(c) and (d)] with various widths.}
\label{fig2}
\end{figure}

However, in the ultrafast spectroscopy measurements, THz ac probe pulses, which usually come from the same source of pump pulses, are employed instead of electric $\delta$-like fields. It seems that a more realistic approach for a probe pulse could be~\cite{Lu12,Rincon14,Fukaya15}
\begin{align}
	 A_{\text{probe}}(t'')=&A_{0,\text{probe}}\,\exp\left[-\left(t''-t\right)^2/2t_{\text{d,probe}}^2\right]\nonumber\\
\times&\cos\left[\omega_{\text{probe}}\left(t''-t\right)\right],
\label{eq:19}
\end{align}
where a Gaussian-like envelope around $t$ is used with $t_{\text{d,probe}}$ to characterize the temporal width of the probe pulse, and $\omega_{\text{probe}}$ is the central frequency. Following the previous conventions, we call the PP method with the Gaussian-like vector potential PP-Gaussian. An illustration of Gaussian pulses with different time widths $t_d$ is shown in Fig.~\ref{fig2}(a).

\begin{figure}
\centering
\includegraphics[width=0.47\textwidth]{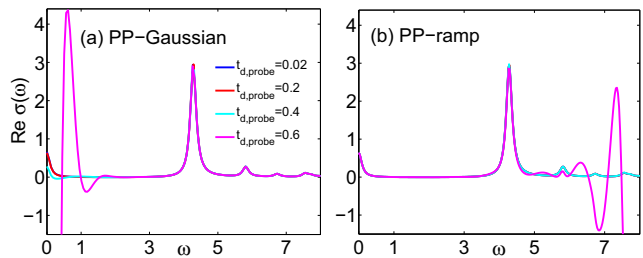}
\caption{(Color online) The calculated $\RE\,\sigma(\omega)$ in equilibrium (zero temperature) for the Hamiltonian~(\ref{eq:1}) with different $t_{\text{d,probe}}$ . Parameters: $L=10$, $U=10$, and $V=4.5$. (a) By PP-Gaussian with $\omega_{\text{probe}}=10$, and $A_{0,\text{probe}}=1.0\times 10^{-6}$. (b) By PP-ramp with $A_{0,\text{step}}=1.0\times 10^{-4}$. Note that the difference is negligible between the results of $t_{\text{d,probe}}=0.2$ (red curves) and those of $t_{\text{d,probe}}=0.02$ (blue curves).}
\label{fig3}
\end{figure}

Let us examine what approximation has to be made in order to justify Eq.~(\ref{eq:15}) in the PP-Gaussian. Suppose that the pulse is activated during $[t,t+\Delta{t}]$, i.e., ${E}_{\text{probe}}(t'')\neq0$ only when $t''\in[t,t+\Delta{t}]$. Consequently, $\vev{j_{\text{probe}}(t')}=0$ when $t'<t$. From Eq.~(\ref{eq:18}) we have
\begin{align}
j_{\text{probe}}(\omega)&=
{\int_{t}^{t+\Delta t}}\sigma(\omega,t'')E_{\text{probe}}(t'')e^{i\omega t''}\,\ud t''\nonumber\\
&\approx\sigma(\omega,t)E_{\text{probe}}(\omega),
\label{eq:20}
\end{align}
which is nothing but Eq.~(\ref{eq:15}). The key approximation lies in the last step of Eq.~(\ref{eq:20}), where $\sigma(\omega,t'')$ is approximated by $\sigma(\omega,t)$. It requires a narrow probe pulse, i.e., $t_{\text{d,probe}}$ should be small enough for nonequilibrium in order to obtain $\sigma(\omega,t)$. It is noted that even in equilibrium, the width of the probe pulse can also influence the resulting $\sigma(\omega)$, as shown in Fig.~\ref{fig3}. In Fig.~\ref{fig3}(a), the zero-temperature optical conductivity $\sigma(\omega)$ in equilibrium for the Hamiltonian~(\ref{eq:1}) on the $L=10$ lattice is calculated by PP-Gaussian under different values of $t_{\text{d,probe}}$. It shows that when $t_{\text{d,probe}}\le 0.2$, the results are converged. The reason is as follows: in PP-Gaussian, the incoming photon frequency is broadened into a Gaussian-like distribution, with the variance of $1/t_d^2$ around $\omega_{\text{probe}}$; in order to cover a large enough $\omega$-regime, a sufficiently small $t_{\text{d,probe}}$ is preferred. This observation explains the large deviation for $t_{\text{d,probe}}=0.6$ at low frequencies when $\omega_{\text{probe}}$ is set to be $10$ in Fig.~\ref{fig3}(a).

To make the information complete, we also investigate the ramping effect in PP-step, where a schematic illustration of $A_{\text{probe}}(t)$ with a finite increase in width (called the PP-ramp) can be found in Fig.~\ref{fig2}(b) (similar to a half-cycle pulse in Ref.~\onlinecite{Tsuji12}). For the Hamiltonian~(\ref{eq:1}) with identical parameters used in the previous PP-Gaussian calculation, we see that the results converge as $t_{\text{d,probe}}\le 0.4$, as shown in Fig.~\ref{fig3}(b). The noticeable deviation for $t_{\text{d,probe}}=0.6$ at high frequencies is due to the exponential decay of $A_{\text{probe}}(\omega)$ in the frequency space with the decay constant proportional to $t_{\text{d,probe}}^2$.

The above discussion tends to suggest that with small enough $t_{\text{d,probe}}$, the results of PP-Gaussian and PP-step would be the same. We have shown that this is actually the case in equilibrium. However, there is a critical difference between them. Mathematically, note that in the PP-Gaussian, unlike the PP-step, $A_{\text{probe}}(t'')\to 0$ as $t''\to\pm\infty$, and consequently, the temporal integration over the electric field gives $\int_{t_i}^{t_f}E(t'')\,\ud t''\approx\int_{-\infty}^{+\infty}E(t'')\, \ud t''=0$, which might be true in experimental situations~\cite{Madsen02}. The induced current $\vev{j_{\text{probe}}(t')}$, as shown in Eq.~(\ref{eq:17}), is produced in PP-Gaussian by a temporally localized perturbation in the range of $[t,t+\Delta t]$ (outside the range, {\em both} the electric field and the vector potential are practically zero). It is therefore reasonable to suggest that when the Gaussian-like pulse is used, only the correlations between the variant $t'$ and (roughly) fixed $t$ (probe time) are taken into account in the calculation of $\sigma(\omega,t)$, which coincides with the feature of VNLR when the temporal correlations are concerned (see Fig.~\ref{fig1}).

It is instructive to consider a limiting case in which we simply put $\omega_{\text{probe}}=0$, and $t_{\text{d,probe}}\to 0$, i.e., using a $\delta$-like vector potential. If we put an electric field corresponding to the $\delta$ vector potential (similar to a monocycle pulse in Ref.~\onlinecite{Tsuji12}) into Eq.~(\ref{eq:5}) by using the response function in NLR, and if we perform a partial integration, we find that the resulting $\vev{j_{\text{probe}}(t')}$ is proportional to $\chi(t',t)$ (here we just focus on the current-current correlation term). Employing its Fourier transformation into Eq.~(\ref{eq:15}), what we finally obtain is merely the expression of the NKF in Eq.~(\ref{eq:10}). The loop between PP-Gaussian and VNLR thus can be closed. From the above analysis, we propose that the PP-step and PP-Gaussian can be identified with the NLR and VNLR, respectively. This proposition will be examined in the next section numerically.



In the following simulations, we set $t_m\sim 200$, $M=30$, $\delta{t}=0.02$, and $\eta=1/\text{L}$, with the lattice size $L=10$. PP-Gaussian and PP-step (as the limiting case of zero-width PP-ramp) are employed, compared with the results of NLR and VNLR. In the PP-step, $A_{0,\text{step}}=1.0\times 10^{-4}$; in the PP-Gaussian, $\text{A}_{\text{0,probe}}\sim10^{-6}$, $\text{t}_{\text{d,probe}}=0.02$, and $\omega_{\text{probe}}=10$. Only the real parts of the optical conductivity are displayed. Despite double temporal evolutions being involved, the performance of the PP methods can match those of (V)NKF in speed, and they are less memory-demanding because no integration is required~\cite{Zala14}.

\section{Results and comparison}\label{sec4}

In this section, working on the 1D half-filled extended Hubbard model~(\ref{eq:1}), we compare the numerical results of the optical conductivity both in and out of equilibrium between various methods.

\begin{figure}
\centering
\includegraphics[width=0.48\textwidth]{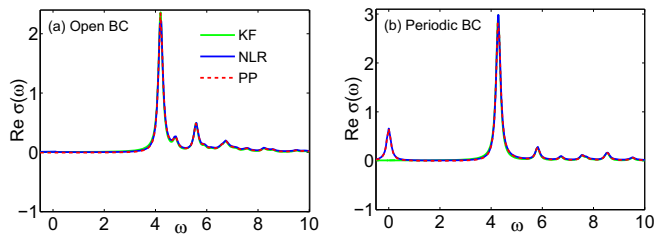}
\caption{(Color online) $\RE\,\sigma(\omega)$ in equilibrium (zero temperature) for the Hamiltonian~(\ref{eq:1}) obtained by the KF, NLR, and PP methods in (a) the open boundary condition (open BC) and (b) the periodic BC. Parameters: $L=10$, $U=10$, and $V=4.5$.}
\label{fig4}
\end{figure}

We first discuss the equilibrium case. Figure~\ref{fig4} shows the results in equilibrium (zero temperature) under open and periodic boundary conditions (BCs), obtained by the KF, NLR, and PP. There are no differences between the results of the PP-step and PP-Gaussian. The NKF, as a generalization of the KF to nonequilibrium, produces data identical to the KF in equilibrium. We can see that the results obtained by the three methods coincide well in the open BC, while in the periodic BC, deviations occur in the vicinity of $\omega=0$. As shown in Fig.~\ref{fig4}(b), an identical peak around zero frequency is produced both by the PP and NLR, while it is absent in the KF result. The small peak is known to originate from the nonzero charge stiffness (Drude weight) in the periodic BC for {\em finite systems}, even when the system is in an insulating phase. The reason is as follows. The charge stiffness $D$ is related to the singularity of $\RE\,\sigma(\omega)$ at $\omega=0$ as~\cite{Castella95,Zotos96}:
\begin{equation}
\RE\,\sigma(\omega)=2{\pi}D\delta(\omega)+\RE\,\sigma_{\text{reg}}(\omega),
\label{eq:21}
\end{equation}
and here
\begin{equation}
D=(1/2L)[\langle\tau\rangle+\RE\,\chi(\omega=0)],
\label{eq:22}
\end{equation}
where $\chi(\omega)$ is defined in Eq.~(\ref{eq:3}). We can see that the Drude weight component comes from the net contribution of the diamagnetic term $\vev{\tau}$ and the current-current correlations at $\omega=0$. For an insulator, $D$ might still remain finite especially in the periodic BC when the system size is small, though it should vanish in the thermodynamic limit. It has been confirmed by the sum rule check (not shown here) that for our model at half-filling with $L=10$, $D$ has a sizable value in the periodic BC, while in the open BC the value is negligibly small. The introduction of the $\eta$ factor in the numerical calculations expands the singularity at $\omega=0$ into a small peak, as shown in Fig.~\ref{fig4}(b). Here we can see that the contribution to the conductivity from the Drude weight can be captured both by the NLR and PP methods, while the KF method, which only takes into account the regular part of the current-current correlations in Eq.~(\ref{eq:2}), fails to produce this feature.

We are now ready to proceed to the nonequilibrium case. As a demonstration, three well-attended situations are considered here. The driving force to nonequilibrium in the first case is a pump pulse with a form similar to Eq.~(\ref{eq:19}) (formally just replace "probe" with "pump"), and we call this situation the Pump case. In the second case, we apply a temporary $\delta$-like electric pulse, which can be simulated by introducing a phase shift $e^{i\phi\theta(t)}$ in the hopping terms of Eq.~(\ref{eq:1})~\cite{Denis12}, where $\theta(t)$ is the Heaviside function. The strength of the pulse is controlled by $\phi$. In our calculations we choose $\phi=\pi$, an extreme case that is known as $\pi$-quench. The third case is the $U$-quench, where a sudden change of $U$ takes place at $t=0$. This situation can be realized, for example, in cold atoms~\cite{Bloch08}.

\begin{figure}
\includegraphics[width=0.47\textwidth]{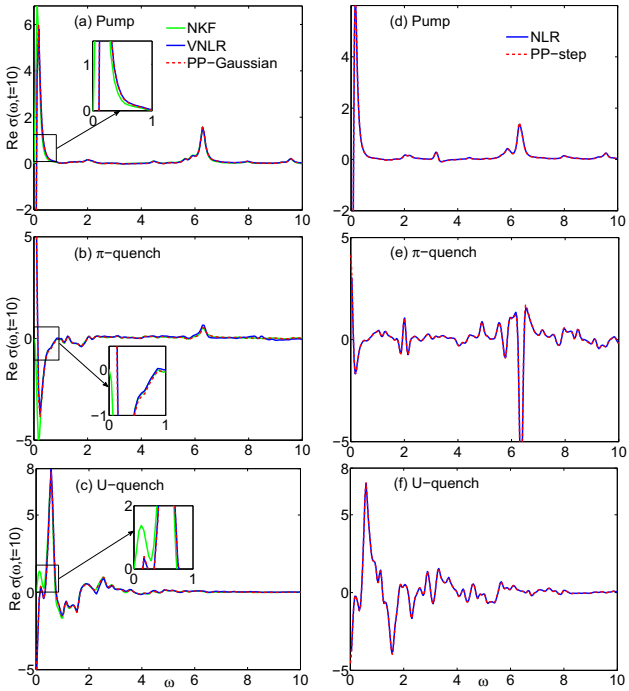}
\caption{(Color online) The time-dependent optical conductivity $\RE\,\sigma(\omega,t)$ at $t=10$ for the Hamiltonian~(\ref{eq:1}) obtained using different methods in the periodic BC, where $t$ is the time delay between the probe and pump (quench) time. In (a) and (d), a pump pulse is applied with parameters $A_{0,\text{pump}}=0.2$, $\omega_{\text{pump}}=6.29$ (resonant frequency), and $t_{\text{d,pump}}=0.5$. In (b) and (e), a $\pi$-quench is applied. In (c) and (f), a $U$-quench with the $U$ value changed from 10 to 4 is applied. The blue solid lines represent the results of the VNLR in the left column ((a), (b), and (c)) and NLR in the right column ((d), (e), and (f)), respectively, while the red dashed lines represent the results of the PP-Gaussian and PP-step in the two columns. The insets in the left column display the details in the vicinity of $\omega=0$ obtained by NKF, VNLR and PP-Gaussian. Model parameters: $L=10$, $U=10$, $V=3$.}
\label{fig5}
\end{figure}

Figure~\ref{fig5} shows the results of $\RE\,\sigma(\omega,t)$ at $t=10$ for these three setups with various methods. First note that the NKF method only includes the regular part of the current-current correlation term (see Eq.~(\ref{eq:7})). As a result, $\RE\,\sigma(\omega,t)$ equals to zero at $\omega=0$ exactly. Away from the vicinity of $\omega=0$, the results of the NKF and VNLR are consistent. This is not surprising since the NKF is just about the regular part of VNLR. More importantly, from Fig.~\ref{fig5}, we can see that in all three cases, including pump and quench, the results between PP-Gaussian and VNLR (left column) and PP-step and NLR (right column) coincide well with each other. Thus the proposition in the previous section of the equivalence among these groups is verified numerically. We make it clear that the NKF and NLR, as being widely used in the dynamic response study, actually correspond to different limits of the probe in the PP method, which either has a Gaussian-like vector potential with a vanishing width, or a steplike one. The former with a tiny width as small as $t_{\text{d,probe}}=0.02$ actually can be regarded as an approximation of the temporal $\delta$-like potential, which has been numerically confirmed in our calculations (not shown here). Though the electric fields are both highly temporally confined in these two cases, one essential difference is the persistence (or temporal locality in the opposite sense) of the vector potential in the subsequent time evolutions, which results in the (non)vanishing of the time integral of the electric field over the probing time. A possible implication of the above observations is that the time-dependent conductivity or other similar transport quantities in nonequilibrium obtained in experimental measurements, might depend on the details of the probe if the features of the measuring procedures are essentially described by the PP method. Theoretical investigations should take into account the relevant experimental setup sufficiently. In the present THz spectroscopy measurements, the duration time of probe pulses is usually hundreds of times larger than the microscopic time unit, which may vindicate the approximation of PP-step or NLR, where the characteristic duration time of the probe is much longer than the time scale of the dynamics in which we are interested~\cite{Nicoletti15}. Conversely, i.e., if the probe time scale is much smaller than the latter, VNLR or PP-Gaussian might be a good starting point for analysis.

It may be worthwhile to make some supplementary comments on the physics in Fig.~\ref{fig5} before closing the section, though it is not the main subject of the present study. In the Pump case, a prominent Drude-like structure of the transient conductivity in the low-frequency regime can be observed, associated with the suppression of the main absorption peak around $\omega=6.3$. It is a typical phenomenon due to photoinduced carriers. More discussions on this issue in terms of this model can be found, e.g., in Ref.~\onlinecite{Lu15}. The $\pi$-quench corresponds to a population inversion in the kinetic energy~\cite{Tsuji12}, where a temporally enhancement of charge mobility is also observed. The $U$-quench is one of the examples of interaction quench~\cite{Eckstein10}. In our case, with $U$ being switched from $10$ to $4$, part of the interaction energy is released into the system with the consequent enhancement of the kinetic energy, which also results in substantial changes in the optical spectrum.

\begin{figure}
\includegraphics[width=0.45\textwidth]{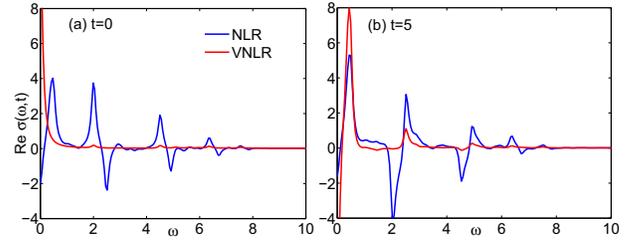}
	\caption{(Color online) The time-dependent optical conductivity $\RE\,\sigma(\omega,t)$ for the Hamiltonian~(\ref{eq:1}) obtained by NLR and VNRL in a $U$-quench case, where $U$ is switched from $0$ to $2$ at $t=0$. The probing time is set at $t=0$ in (a) and $t=5$ in (b), respectively. Model parameters: $L=10$ and $V=0$.}
	\label{fig6}
\end{figure}

Additionally, as shown in Fig.~\ref{fig5}, a similar development of $\sigma(\omega,t)$ between the left and right columns can still be recognized regardless of the notable difference in quenches. This is consistent with the previous statement in Sec.~\ref{sec2} that the VNLR is a kind of partial summation on the full temporal correlations in NLR. In the Pump case, the results of VNLR and NLR are closer to each other~\cite{Lu15} compared with the cases of quenches, which might be due to the fact that the Hamiltonian only changes temporarily with the application of the pump pulse, after which it goes back to its earlier form; in both $\pi$-quench and $U$-quench, the Hamiltonian undergoes a substantial change after the quench time. Numerical checks in the periodic BC up to $L=14$ and the open BC in $L=10$ show similar results. Thus we tend to conclude that the deviation between VNLR and NLR, which is due to different temporal correlation counting, persists with the increase of the system size. Further, we note that the deviation of VNLR from NLR can be quite pronounced if the probing time is set to be close to the pump or quench moment before the relaxation fully commences. This is because, in contrast with NLR, VNLR only considers the temporal correlations starting from the probing time $t$ (see Fig.~\ref{fig1}). Figure~\ref{fig6} shows $\RE\,\sigma(\omega,t)$ for the 1D half-filled Hubbard model ($V=0$ for the Hamiltonian~(\ref{eq:1})) in a typical $U$-quench case, where the interaction $U$ is switched from $0$ to $2$ at $t=0$. The probing time is set to be $t=0$ and $t=5$ in (a) and (b), respectively. Note that at $U=0$, the ground state of the system is in a metallic state, and with any finite $U$ a charge gap is opened and the system becomes insulating. In Fig.~\ref{fig6}(a), we can observe that the resulting $\RE\,\sigma(\omega,0)$ from VNLR shows a strong reminiscence of the initial metallic state with a prominent Drude-like peak appearing near zero frequency, while the result from NLR, which takes into account more information about the subsequent nonequilibrium evolution, not only shows the depression of this peak, but it also produces additional structures in the higher-frequency regime. However, when the probing time is set at $t=5$, by which time the relaxation between the kinetic energy and the interaction energy is largely finished, the results from VNLR and NLR bear more qualitative similarities.

\section{Conclusion}\label{sec5}

In this paper, we introduce a numerical method to calculate the time-dependent optical conductivity based on a simulation of the pump-probe experimental setup. According to the property of the probe pulse, two different approaches are discussed, i.e., PP-step and PP-Gaussian, using either a step-like or a Gaussian-like probing vector potential. We find that the two approaches can be identified with nonequilibrium linear-response theory and one of its variants, respectively. Thus the connections between these various methods and their differences are clarified. Numerical verification is given by employing the method systematically in the 1D half-filled extended Hubbard model, both in and out of equilibrium. The latter includes three well-attended cases: the pump, $\pi$-quench, and $U$-quench. In the numerical results, the probe-pulse dependence in nonequilibrium is especially significant in the quenches. We suggest that the nature of the probe pulses should be taken into account in the analysis of ultrafast THz spectroscopy measurements.


\begin{acknowledgments}
The authors thank Janez Bon\v{c}a, Denis Gole\v{z} and Dirk Manske for helpful discussions.
H.L. acknowledges support from the National Natural Science Foundation of China (NSFC) (Grants Nos. 11325417, 11474136, and 11174115), and the open project of State Key Laboratory of Theoretical Physics (SKLTP) in ITP, CAS.
T.T. acknowledges support by the Grant-in-Aid for Scientific Research (26287079) from MEXT and the Strategic Programs for Innovative Research (SPIRE) (hp140215, hp150211), the Computational Materials Science Initiative (CMSI).
\end{acknowledgments}


%

\end{document}